\documentstyle[osa,manuscript]{revtex}

\begin{document}
\title{Interferometry of hyper-Rayleigh scattering\\
by inhomogeneous thin films}
\author{A.A. Fedyanin, N.V. Didenko, N.E. Sherstyuk,  A.A. Nikulin,
O.A. Aktsipetrov\footnote{E-mail address: aktsip@astral.ilc.msu.su, Web:  
http://kali.ilc.msu.su}}
\address{Department of Physics, Moscow State University, Moscow 119899, Russia}
\date{\today}
\maketitle

\begin{abstract}
The use of specific symmetry properties of the optical second-harmonic 
generation (the {\it s},{\it s}-exclusion rule) has allowed us to observe 
high-contrast hyper-Rayleigh
interference patterns in a completely diffuse light - an effect having no
analog in case of linear (Rayleigh) scattering. \ 
\end{abstract}

\vspace{5mm} \noindent

\newpage

In the past few years the versatility of optical second-harmonic generation
(SHG) \cite{1} as a probe for studying solid state nanostructures and
low-dimensional systems has been embodied in experimental methods of SHG
interferometry (phase measurements) \cite{2} and hyper-Rayleigh
scattering (HRS) \cite{3}. These methods exploit opposite features of the 
second-harmonic (SH) radiation generated by an object being studied: the former
implies coherence of the SH radiation, whereas the latter deals with
incoherent SHG originating from spatial fluctuations of the optical
parameters in a randomly inhomogeneous system. Therefore the very idea of
the HRS interferometry may seem contradictory, which
is, however, disproved by a more detailed analysis of a conventional
experimental scheme of SHG interferometry. In this scheme shown in Fig. \ref
{one}, panel 1, the interference pattern is formed by the SH wave from a sample
studied and that from another object having well-defined optical
characteristics and used as a reference. The phase shift between the two
waves results from the dispersion of the refraction index of air $n$, which
yields the following dependence of the total SHG intensity $I_{2\omega }$ on
the reference translation:

\begin{equation}
I_{2\omega }(x_{\text{{\rm R}}})=I_{2\omega }^{\text{{\rm S}}}+I_{2\omega }^{%
\text{{\rm R}}}+2\alpha \sqrt{I_{2\omega }^{\text{{\rm S}}}I_{2\omega }^{%
\text{{\rm R}}}}\cos \left[ \Phi _{0}+K(x_{\text{{\rm R}}}-x_{\text{{\rm S}}%
})\right] \text{,}  \label{eq1}
\end{equation}
where $I_{2\omega }^{\text{{\rm S}}}$ and $I_{2\omega }^{\text{{\rm R}}}$
are the intensities of the SH waves from the sample and the reference,
respectively, $\alpha $ is the mutual coherence factor of the two waves, $x_{%
\text{{\rm S}}}$ and $x_{\text{{\rm R}}}$ are the coordinates of the sample
and the reference, respectively, measured in the reference translation
direction, $\Phi _{0}$ is position-independent part of the relative phase
shift between the two waves, $K=2\omega (n(2\omega )-n(\omega ))/c$ and $%
\omega $ is the fundamental radiation frequency. The essential point is that
the SH wave from the reference is then reflected from the sample. It is the
feature of the interferometry scheme that allows us to observe HRS
interference patterns. A randomly inhomogeneous sample is a source of both
HRS of the fundamental radiation and Rayleigh
scattering (RS) of the SH wave generated by the reference and then linearly
reflected by the sample. In other words, the SH wave from the reference,
after being reflected from the sample, becomes 'modulated' by the spatial
fluctuations of the linear-optical properties of the sample. Therefore the
interfering SH fields from the reference and the sample will both have
incoherent components that are {\it mutually} coherent (statistically
dependent) because of their common statistical origin: the spatial
fluctuations of the optical parameters of the sample. This modifies Eq.(\ref
{eq1}) for the resulting interference pattern as follows:

\begin{eqnarray}
I_{2\omega }(x_{\text{{\rm R}}})=I_{2\omega }^{\text{{\rm S}}}+I_{2\omega
}^{\text{{\rm R}}}+2\bar{\alpha}\sqrt{\bar{I}_{2\omega }^{\text{{\rm S}}}%
\bar{I}_{2\omega }^{\text{{\rm R}}}}\cos [\bar{\Phi}+Kx_{\text{{\rm R}}}]+
2\widetilde{\alpha }\sqrt{\widetilde{I}_{2\omega }^{\text{{\rm S}}}%
\widetilde{I}_{2\omega }^{\text{{\rm R}}}}\cos [\widetilde{\Phi }+
Kx_{\text{{\rm R}}}],  \label{eq2}
\end{eqnarray}
where $I_{2\omega }^{\text{{\rm S}}}=\bar{I}_{2\omega }^{\text{{\rm S}}}+%
\tilde{I}_{2\omega }^{\text{{\rm S}}}$ , $I_{2\omega }^{\text{{\rm R}}}=%
\bar{I}_{2\omega }^{\text{{\rm R}}}+\tilde{I}_{2\omega }^{\text{{\rm R}}}$
and the bar (tilde) over a quantity denotes a coherent (incoherent)
component. For the {\it s}-in,{\it s}-out wave-polarization combination
the coherent component $\bar{I}_{2\omega }^{\text{{\rm S}}}$ 
vanishes due to the {\it s,s}-exclusion rule\cite{4}, and
solely an HRS interference pattern is observed.

In this paper we study two types of inhomogeneous solid films: those of
purple membranes of
bacteriorhodopsin (bR) and polycrystalline ${\rm %
Pb_{x}(Zr_{0.53}Ti_{0.47})O_{3}}$ (PZT) ferroelectric films. bR films with a
thickness of about 1000 nm were prepared by
the method described in Ref. [\cite{5}]. PZT texture films were fabricated
by the sol-gel technique\cite{6} and then annealed at 650$^{0}$C.
For SHG experiments the output of a Q - switched YAG:Nd$^{3+}$ laser at 1064
nm is used as a fundamental radiation with the pulse duration, repetiton
rate and intensity being 15 ns, 12.5 Hz and 1 MW/cm$^{2}$, respectively
The {\it s}-polarized fundamental wave irradiates
the sample at an angle of
incidence $\theta _{0}=45^{0}$. Having passed through a set of color
filters,  the {\it s}-polarized SH component is then detected
by a PMT. The interference patterns $I_{2\omega }(x_{{\rm R}})$ are 
obtained by
translating the reference along the laser beam. The reference is a 2
mm-thick plate of fused quartz coated by a 30 nm-thick indium-tin oxide
film. Declining the reference by (0$\div$5)$^0$ from the
position perpendicular to the laser beam (see panel 1 in Fig. \ref{one})
affects $I_{2 \omega}^{\rm R}$ but not the
intensity of the fundamental radiation transmitted through the reference.
This allows us to balance the values of $I_{2\omega }^{{\rm S}}$ and 
$I_{2\omega }^{{\rm R}}$ and thus to provide a maximal 
interference-pattern contrast attained at 
$I_{2\omega }^{{\rm S}}\approx I_{2\omega }^{{\rm R}}$.
The values of $I_{2\omega }^{{\rm S}}$ and 
$I_{2\omega }^{{\rm R}}$ are measured separately by inserting
an appropriate filter (infrared or green, respectively) in between
the reference and sample.
Fig. \ref{one} shows
the interference patterns for bR (panel 2) and PZT (panel 3) films.
Well-pronounced oscillations of the total SHG intensity are observed.

Three additional optical experiments are performed for more thorough 
interpretation of the HRS interferometry data. First, the dependences of
the SH intensity $I_{2\omega }$ on the azimuthal angle $\psi $ of the sample
rotation about the axis normal to the film plane were measured and are
presented in insets in Figs. \ref{two}, \ref{three}. For the bR film
the dependence $I_{2\omega }(\psi )$ is practically isotropic within the
error bars, whereas for the PZT film a pronounced anisotropy with one-fold
symmetry occurs\cite{7}. The interference pattern shown in Fig. \ref{one}, panel 3
was measured in the minimum of this dependence. Second, the dependences of
the SH intensity on the polar scattering angle $\theta $ (HRS indicatrix)
were measured and are shown in Figs. \ref{two}, \ref{three}. 
For both types of films the HRS indicatrices have maxima at $\theta
=\theta _{0}$, but no SH signal jump is observed in the vicinity of the
specular direction. This proves the absence of the regular SHG contribution in
the specular direction for the {\it s}-in, {\it s}-out polarization
combination at the values of the azumuthal angle chosen for the HRS
interferometry measurements. Third, linear, or Rayleigh, scattering 
indicatrices were measured at the SH wavelength and are shown in 
Figs. \ref{two}, \ref{three}. The RS indicatrices, in
contrast with the HRS ones, have sharp specular peaks corresponding to the
coherent reflection of light.

Both the HRS indicatrix and that of RS with the coherent-reflection peak
subtracted can be approximated by a superposition of two Gaussian peaks:

\begin{eqnarray}
I_{\sigma }(\theta )=I_{\sigma }T(\theta )
\left( {\exp }\left\{ -\left[ k(\theta) l_{\sigma }\right] ^{2}\right\} 
{+\gamma }_{\sigma }{\exp }\left\{ -\left[ k(\theta) L_{\sigma }\right] 
^{2}\right\} \right) \text{,}  
\label{eq8}
\end{eqnarray}
where $k(\theta)=2\omega \left( {\sin \theta -\sin \theta _{0}} \right)/c$, 
$\sigma =1$, $2$, the subscripts $1$ and $2$ denote the RS and HRS
indicatrices, respectively, $T(\theta )$ is a known function being the
transmission coefficient of the film-air interface at the SH wavelength with
the dielectric constant of the film as an adjustable parameter, $I_{\sigma }$%
, ${\gamma }_{\sigma }$, $l_{\sigma }$, $L_{\sigma }$ are adjustable
parameters. The quantities $l_{\sigma }$ and $L_{\sigma }$ are correlation
lengths characterizing the in-plane scales of spatial fluctuations of
optical parameters in the films. The fitting to the experimental data yields 
$l_{1}^{{\rm bR}}\simeq l_{2}^{{\rm bR}}\simeq $ 200 nm, $L_{1}^{{\rm bR}%
}\simeq L_{2}^{{\rm bR}}\simeq $ 1000 nm, $l_{1}^{{\rm PZT}}\simeq l_{2}^{%
{\rm PZT}}\simeq $ 170 nm, $L_{1}^{{\rm PZT}}\simeq L_{2}^{{\rm PZT}}\simeq $
2000 nm. Thus, for each film type RS and HRS have
two common fluctuating sources that are statistically independent and
characterized by Gaussian correlation functions with quite different
correlation lengths $l$ and $L$. One of the sources is likely to be the 
inhomogeneities in bR and PZT films: 
membrane aggregates and microcrystallites, respectively. 
The other one presumably is the film interfacial roughness.

Since the interference patterns are measured under conditions that provide
vanishing of the coherent SH signal from the sample, Eq.(\ref{eq2}) takes
the form:

\begin{equation}
I_{2\omega }^{s-s}(x_{{\rm R}})=\widetilde{I}_{2\omega }^{{\rm S}}+I_{2\omega }^{{\rm R}%
}+2\widetilde{\alpha }\sqrt{\widetilde{I}_{2\omega }^{{\rm S}}\widetilde{I}%
_{2\omega }^{{\rm R}}}\cos [Kx_{{\rm R}}+\widetilde{\Phi} ]\text{,}  \label{eq3}
\end{equation}
where $\widetilde{I}_{2\omega }^{{\rm S}}$ and $I_{2\omega }^{{\rm R}}$
are measured quantities, $\widetilde{\alpha }$ is an adjustable parameter and
 $\widetilde{I}_{2\omega }^{{\rm R}}=I_1(\theta_0)$ is the value estimated from
the maximun of RS indicatrix. In order to 
elucidate the statistical meaning of $\widetilde{\alpha }$ we
relate $\widetilde{I}_{2\omega }^{{\rm R}}$ and $\widetilde{I}%
_{2\omega }^{{\rm S}}$ to the constituents of the
polarization induced in the sample: 
$\widetilde{I}_{2\omega }^{{\rm S}}\propto \langle |\widetilde{{\bf P}}%
_{2\omega }^{{\rm S}}|^{2}\rangle$, $\widetilde{I}_{2\omega }^{{\rm R}%
}\propto \langle |\widetilde{{\bf P}}_{2\omega }^{R}|^{2}\rangle$.
Here $\widetilde{{\bf P}}_{2\omega }^{{\rm S}}$ is the total polarization
(i.e. the sum of quadratic and linear polarizations) induced in the sample
at the SH frequency by the fundamental radiation, $\widetilde{{\bf P}}%
_{2\omega }^{{\rm R}}={\bf P}_{2\omega }^{{\rm R}}-\langle {\bf P}_{2\omega
}^{{\rm R}}\rangle $, ${\bf P}_{2\omega }^{{\rm R}}$ is the linear
polarization induced in the sample upon reflection of the SH wave generated
by the reference, and angular brackets denote averaging over spatial
fluctuations. Hence $\widetilde{\alpha }$ is defined as follows:

\begin{equation}
\widetilde{\alpha }={\rm Re}\langle (\widetilde{{\bf P}}_{2\omega }^{%
{\rm R}})^{\ast }\cdot \widetilde{{\bf P}}_{2\omega }^{{\rm S}}\rangle 
/\sqrt{\langle |\widetilde{{\bf P}}_{2\omega }^{{\rm R}}|^{2}\rangle \langle |%
\widetilde{{\bf P}}_{2\omega }^{{\rm S}}|^{2}\rangle }\text{.}  
\label{eq7}
\end{equation}

Thus the key point of our explanation of the mutual coherence occurring for
the SH wave from reference and the diffuse SH signal from the sample is the
fact that the SH radiation from the reference is subsequently reflected from
the sample, which leads to statistical dependence of the signals.

According to the data on the RS\ and HRS indicatrices for both bR and PZT
films, each HRS interference pattern can be decomposed into superposition of
two interference patterns originating from the two statistically
independent sources with correlation lengths $l$ and $L$. Taking into
account that, according to Eq.(\ref{eq8}), $\widetilde{I}_{2\omega }^{{\rm R%
}}=I_{1}(\theta _{0})=I_{1}\left( 1+{\gamma }_{1}\right) $, $\widetilde{I}%
_{2\omega }^{{\rm S}}=I_{2}(\theta _{0})=I_{2}\left( 1+{\gamma }_{2}\right) $%
, we re-write Eq. (\ref{eq3}) as follows:

\begin{equation}
I_{2\omega }^{s-s}(x_{{\rm R}})=\bar{I}_{2\omega }^{\text{{\rm R}}%
}+I_{1}\left( 1+{\gamma }_{1}\right) +I_{2}\left( 1+{\gamma }_{2}\right) +2a%
\sqrt{I_{1}I_{2}}\left( \cos [Kx_{{\rm R}}+\widetilde{\Phi}_l ]+
\sqrt{{\gamma }_{1}{\gamma}_{2}}\cos [Kx_{{\rm R}}+\widetilde{\Phi}
_L ]\right) \text{,}  \label{eq9}
\end{equation}
where $0\leq a\leq 1$, the parameters $I_{1,2}$ and ${\gamma }_{1,2}$ have
the same values as in Eq. (\ref{eq8}), whereas the parameters $a$, 
$\widetilde{\Phi}_l$ and $\widetilde{\Phi}_L$ are treated as 
adjustable ones. The phenomenological parameter $a$ 
is introduced to take into account all external factors (such as
incomplete coherence of the fundamental radiation) that reduce mutual
coherence of the HRS and RS signals.
Fitting the experimental data for interference patterns with Eq. (\ref{eq9}%
), we obtain $a^{{\rm bR}}=$ 0.91, $a^{{\rm PZT}}=$ 0.89, 
$\widetilde{\Phi} _{l}^{\rm bR}-\widetilde{\Phi} _{L}^{\rm bR}=1.7$ rad and 
$\widetilde{\Phi} _{l}^{\rm PZT}-\widetilde{\Phi} _{L}^{\rm PZT}=3.1$ rad.

To summarize, interferometry of incoherent second-harmonic generation, or
hyper-Rayleigh scattering, by inhomogeneous films of bacteriorhodopsin and
ferroelectric ceramics has been carried out. High-contrast interference
patterns are observed. This is interpreted as a result of mutual coherence
of the second-harmonic radiation from the film proper and the reference
second-harmonic wave incoherently reflected by the film.

We are thankful to A.S. Sigov, E.D. Mishina for fruitful discussions,
K.A. Vorotilov for preparing PZT samples, and  E.P. Lukashev for 
preparing bR samples. This work was supported by RFBR grants 
97-02-17919, 97-02-17923, 96-15-96420 and 96-15-96476,
INTAS-93 grant 0370(ext), INTAS-RFBR-95 grant 722, 
INTAS grant YSF-9, ISSEP grant d98-701, Programs "Center of 
Fundamental Optics and  Spectroscopy" and 
"Universities of Russia" grant 1412.

\begin{figure}[tbp]
\caption{Panel 1: the schematic of HRS interferometry. Panels 2 and 3: SHG
interference patterns measured in the s-in, s-out wave-polarization
combination and in the absence of the specular SH component for bR and PZT
films, respectively. Solid curves: the dependences given by by Eq.(\ref{eq3}%
) with $\widetilde{\alpha }^{\text{{\rm bR}}}$=0.51, 
$\widetilde{\alpha }^{\text{{\rm PZT}}}$=0.73.
Dashed curves: partial interference patterns, referred to as patterns ''$l$%
'' and ''$L$'', from the RS/HRS sources with the correlation lengths $l$ and 
$L$, respectively. The inset: the schematic of the
pattern formation.}
\label{one}
\end{figure}

\begin{figure}[tbp]
\caption{RS (open circles) and HRS (solid circles) indicatrices for bR film. Solid
curves: the dependences given by Eq.(\ref{eq8}) with $\gamma _{1}^{{\rm bR}%
}=3.0$, $\gamma _{2}^{{\rm bR}}=0.34$. Dashed lines show characteristic
levels of HRS and RS intensities. The inset: the SHG azimuthal dependence
and its approximation by the zeroth Fourier component.}
\label{two}
\end{figure}

\begin{figure}[tbp]
\caption{RS (open circles) and HRS (solid circles) indicatrices for PZT film. Solid
curves: the dependences given by Eq.(\ref{eq8}) with $\gamma _{1}^{{\rm PZT}%
}=293.0$, $\gamma _{2}^{{\rm PZT}}=2.9$. Dashed lines show characteristic
levels of HRS and RS intensities. Left inset: the SHG azimuthal dependence
and its approximation by first two Fourier components. An arrow indicates
the angle $\psi _{0}$ at which the SHG interference patterns are measured. Right
inset: the RS indicatrix in the logarithmic scale.}
\label{three}
\end{figure}

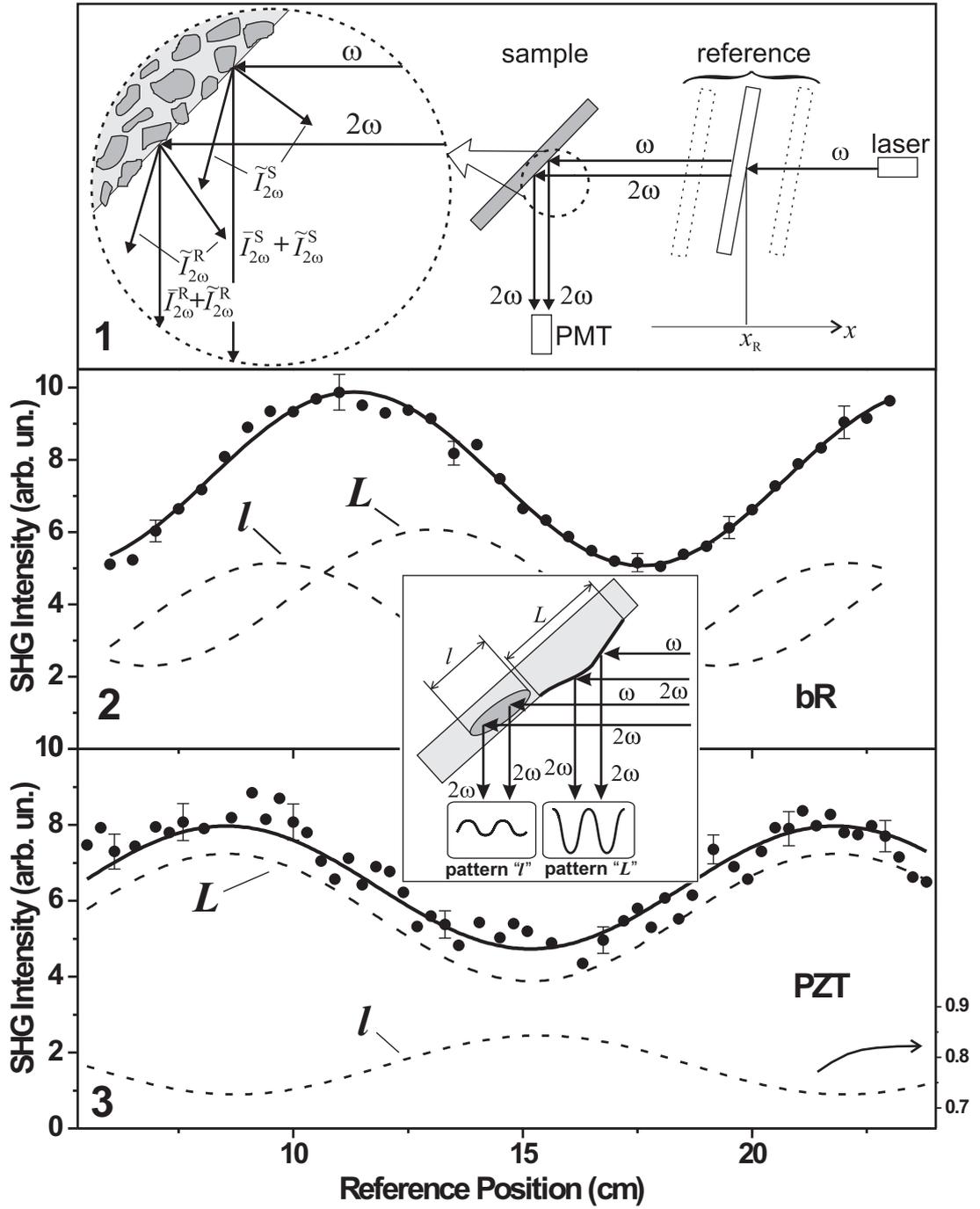
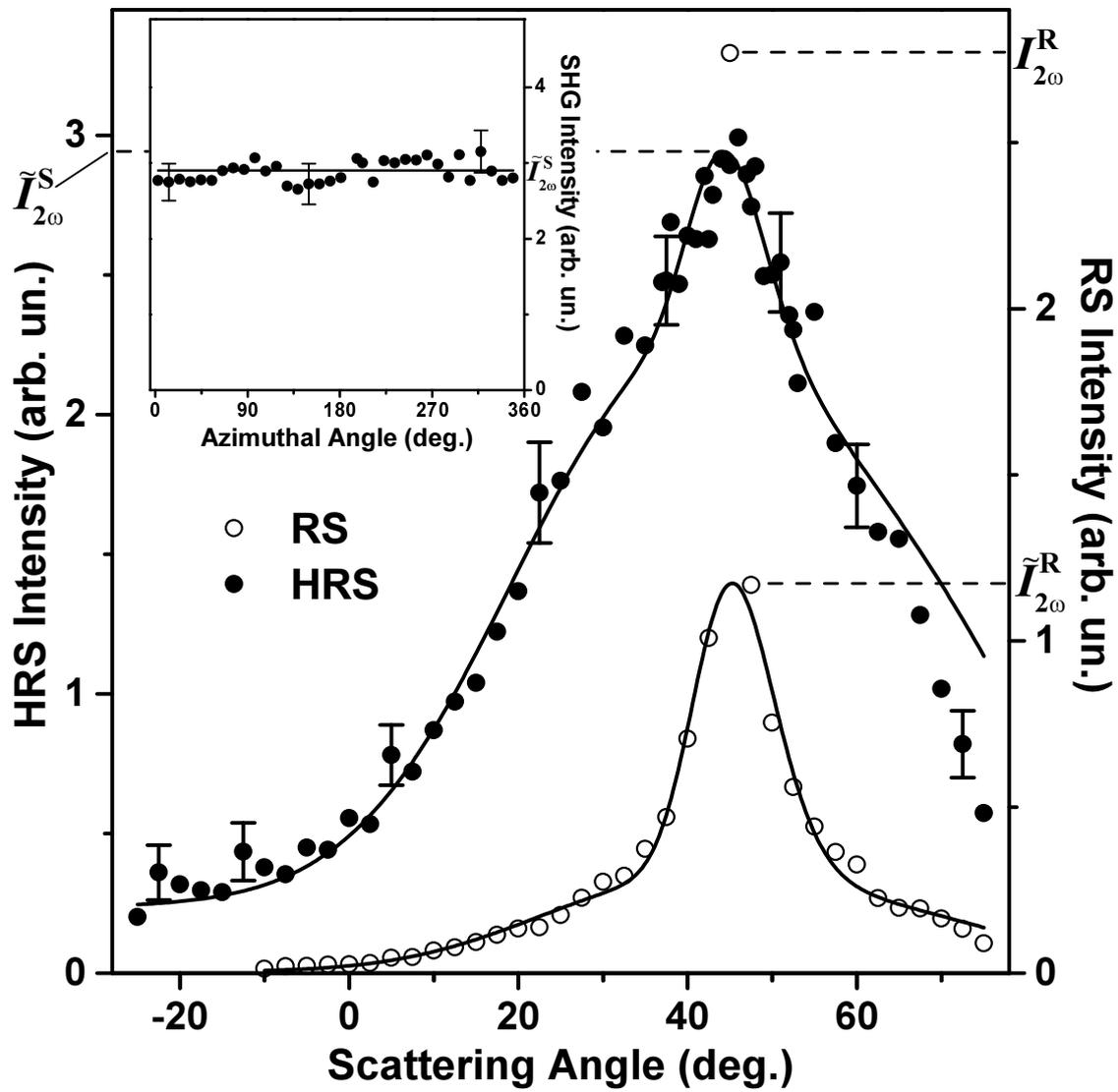
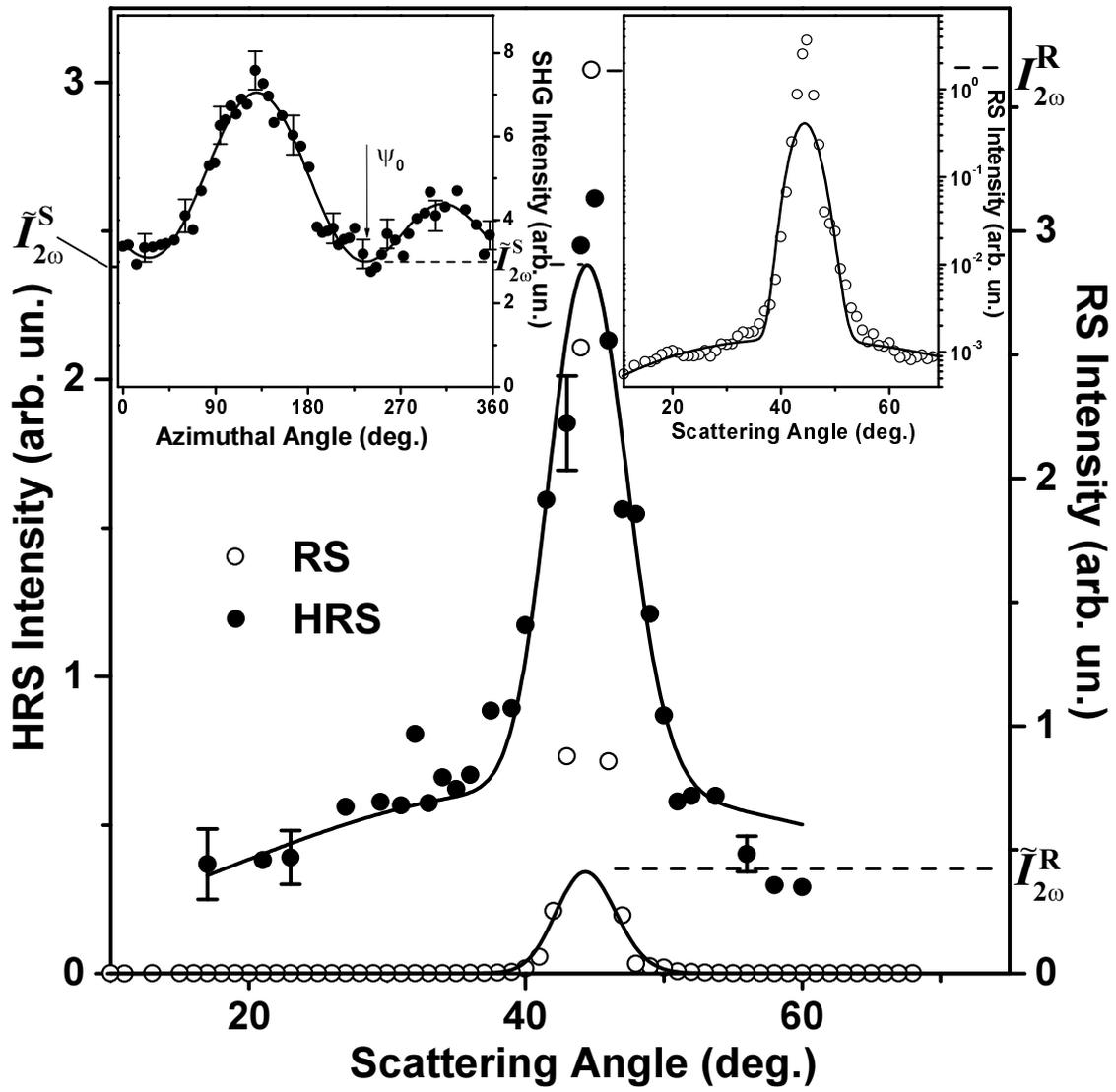
\end{document}